\documentclass[12pt]{iopart}
\pdfoutput=1
\usepackage[english]{babel}

\usepackage{graphicx}
\usepackage{color}					
\usepackage{bm}      				
\usepackage{chemarrow}				
\usepackage{booktabs}				
\usepackage{array}					
\usepackage{slashed}  				
\usepackage[hidelinks]{hyperref}	
\usepackage{tikz}					
\usepackage{verbatim}
\usepackage{notes2bib}				
\bibnotesetup{note-name =,use-sort-key = false}

\usepackage{amsthm}
\usepackage{amsfonts}
\usepackage{amssymb}
\usepackage{array}
\usepackage{bbold}
\usepackage{dsfont}
\usepackage{xcolor}
\definecolor{mypink}{RGB}{237, 152, 152}
\definecolor{myblue}{RGB}{161, 182, 242}
\usepackage{hyperref}
\hypersetup{colorlinks=true, citecolor=blue, linkcolor=blue,urlcolor= blue}

\expandafter\let\csname equation*\endcsname=\relax
\expandafter\let\csname endequation*\endcsname=\relax
\usepackage{amsmath}

\newcommand{\bea}{\begin{eqnarray}}
\newcommand{\eea}{\end{eqnarray}}
\newcommand{\beq}{\begin{equation}}
\newcommand{\eeq}{\end{equation}}
\newcommand{\bit}{\begin{itemize}}
\newcommand{\eit}{\end{itemize}}

\renewcommand{\r}{\right \rangle}
\renewcommand{\l}{\left \langle}
\newcommand{\exy}{(xy,\nu)}

\DeclareMathAlphabet{\mathpzc}{OT1}{pzc}{m}{it}

\newcommand{\1}{\mathpzc{1}}
\newcommand{\2}{\mathpzc{2}}

\newcommand{\X}{{\scriptscriptstyle{X}}}				

\newcommand{\I}{\textsc{i}}
\newcommand{\II}{\textsc{ii}}

\begin{document}
\title{Dynamical equivalence classes for Markov jump processes}
\author{Gatien Verley}
\address{Universit\'e Paris-Saclay, CNRS/IN2P3, IJCLab, 91405 Orsay, France}
\date{\today}

\begin{abstract}
	Two different Markov jump processes driven out of equilibrium by constant thermodynamic forces may have identical current fluctuations in the stationary state. The concept of dynamical equivalence classes emerges from this statement as proposed by Andrieux for discrete-time Markov chains on simple graphs. We define dynamical equivalence classes in the context of continuous-time Markov chains on multigraphs using the symmetric part of the rate matrices that define the dynamics. The freedom on the skew-symmetric part is at the core of the freedom inside a dynamical equivalence class. It arises from different splittings of the thermodynamic forces onto the system's transitions. 
\end{abstract}
\maketitle

\section{Introduction}

A current is intuitively a flow of a physical quantity, such as energy, matter or charges, that changes sign upon time reversal. Currents exist in different spaces. In an abstract state space on which are defined state probabilities, one may consider probability currents between two states. If the state space is the real space, like in the diffusion of a single colloidal particle, a probability current can be thought (with an ensemble viewpoint) as a matter current at the microscopic scale. From such local currents between states or given locations in space, one can define more global currents. For instance across a system-reservoir interface, the sum of local currents for a given physical quantity leads to a macroscopic current through the interface. 

Conservation laws strongly constrains currents at all scales, e.g. probability conservation constrains probability currents while conservation of energy, matter or charges constrains physical currents exchanged with the environment. Each conservation law creates a relation between the currents of a system, reducing by one the number of currents needed to determine all currents in the stationary state \cite{Polettini2014_vol141}. For instance, in the framework of electric circuits, Kirchhoff currents law at each node creates a constraint between currents in each branch of the circuit. Similarly, in the framework of stochastic thermodynamics on a graph, currents along fundamental cycles are enough to determine the currents along any edge of the graph \cite{Schnakenberg1976_vol48}. In the same idea, a subset of physical currents, the so-called fundamental currents \cite{Polettini2016_vol94}, are enough to determine all physical currents exchanged with the reservoirs. Hence, it exists a freedom in the choice of the currents to be measured so that any other current can be determined thanks to the conservation laws. 

In most applications, such as conversion or transport processes, predictions of physical currents must be done in consistency with  conservation laws. Those predictions are made on average, like in hydrodynamic theories \cite{Notes_Beijeren2011} or with statistical distribution in stochastic thermodynamics \cite{Broeck2013_vol,Book_Sekimoto2010}. The most convenient way to determine the statistics of currents is to compute the Laplace conjugate of their probability distribution, namely the moment generating function. However, the aforementioned freedom on the observation of currents translates here in a freedom in the specific form of the stochastic variable whose moments are studied \cite{Wachtel2015_vol92}. For instance, one may observe energy exchanges with a heat reservoir for every transition between different states. Alternatively, one may look at specific and well chosen transitions only and assume that energy is exchanged exclusively during those transitions, but in amounts that compensate the energy that should have been exchanged during other transitions. In short, different splittings of physical currents into local currents are possible, all leading to the same statistics of physical currents. This freedom in the form of the studied stochastic observable corresponds to a gauge freedom for the dynamics \cite{Wachtel2015_vol92, Garrahan2016_vol2016}. As long as one observes over a long time a system in its stationary state, such changes of dynamics have no effects on the trajectory ensemble and thus on the statistics of physical currents. 

This possibility of different systems sharing the same current statistics raises the question of equivalence between the dynamics of nonequilibrium systems. This question has been explored extensively in the last decades for Markov processes in steady states. 
Transition rates including the least possible information on the dynamics, given the measure of some physical currents, have been derived using the maximum entropy principle \cite{Evans2005_vol38}. 
For processes conditioned on observables like physical currents, either by filtrating the trajectory ensemble or by exponentially tilting it (with or without Doob rectification), several asymptotically equivalent dynamics have been obtained \cite{Popkov2010_vol2010, Chetrite2013_vol111, Budini2014_vol2014, Chetrite2015_vol16, Chetrite2015_vol2015, Chabane_2020}. {In these works, the notion of non-equilibrium equivalence may refer to path ensemble equivalence (dynamical equivalence) or to equivalence for the typical value of the observable (concentration equivalence) \cite{Touchette2017_vola}. 
The former equivalence implies the latter that is hence less restrictive. As we detail in \ref{appendix-MeanEquiv}, concentration equivalence can be produced by two kinds of modifications of the rate matrix in the case of jump processes:} First, the stationary state probability and the probability currents are not affected by a \emph{symmetric} modification of the jump probability \cite{Zia2007_vol}, i.e. same modification for each forward and backward jump probability. Second, the probability currents are not affected as well by a modification of the mean time spent in each of the system states \cite{Polettini2012_vol97}. Nonetheless, the stationary probability has changed in this second case, e.g. it increases for states with bigger occupancy time.
{The dynamical equivalence is physically more relevant motivating us to find practical conditions producing it, as Andrieux did for (discrete-time) Markov chains on simple graphs \cite{Andrieux2012_vola, Andrieux2012_vol}. Indeed, defining nonequilibrium thermodynamic potentials is sometime possible thanks to dynamical equivalence \cite{Verley2016_vol93}. In this case, one can connect the rare fluctuations of an equilibrium process with the typical fluctuations of the \emph{same} process put in stationary nonequilibrium by external forces or multiple reservoirs at different intensive parameters.}

Although modifying dynamics while keeping comparable statistics of physical currents has been thoroughly studied, a definition of \emph{dynamical equivalence classes} is still missing for Markov jump processes on multigraph, in continuous time, and driven out of equilibrium by stationary forces or competing reservoirs. We aim to provide such an operational definition in this article.
In section \ref{sec:Notations}, we review standard techniques to compute cumulant generating functions based on the spectrum of tilted rate matrices. 
In section \ref{similarity}, we find a similarity transformation connecting a (tilted) rate matrix to its (tilted) symmetrized form called \emph{root matrix}. 
Based on this result, we define in section \ref{equivalence} a non-equilibrium equivalence class by the ensemble of rate matrices with the same root matrix chosen as the representative of the class. Thanks to the similarity transformation with the root matrix, the identity of the spectrum of the matrices in the equivalence class is guaranteed and so do the fluctuations of physical currents.
We end this article by illustrating the equivalence of fundamental currents fluctuations on a solvable model of molecular motor.

Finally, from the concept of equivalence class we see that thermodynamic forces may appear in many ways in the dynamics with no consequences on the stationary state. 
In \ref{edge-cycle-wise}, the explicit relation between equivalent dynamics is used to explain the freedom on the specific form of the stochastic variable representing a physical current when computing its cumulants~\cite{Wachtel2015_vol92}.

\section{Fluctuations of fundamental currents for systems in non-equilibrium stationary state}
\label{sec:Notations}

We study a system with a finite number of states, modeled by a Markov jump process in continuous time and put out of equilibrium by thermodynamic forces gathered into vector $ \bm{b} = (b_{\I}, b_{\II}, \dots, b_{\X}, \dots) $. These forces are generated by several reservoirs assumed to be always in thermodynamic equilibrium (e.g. thermostat, chemostat) with different intensive parameters. To shorten notations, we denote half of the thermodynamic forces by $\bm{f} = \bm{b}/2$, and talk of $\bm{f}$ as a force by abuse of language.

Each reservoirs exchanges physical quantities (e.g. energy, matter) with the system enabling its change of state, denoted $x$, $y$ or $z$, via channels labeled by $\nu = \1, \2, \mathpzc{3}, \dots$ Hence, $\nu$ labels both a channel and the reservoir involved in the transition via this channel. In the framework of graph theory, a change of system state via a channel is represented by an edge. We label edges by $e$ in general or by $\exy$ for a specific transition from state $y$ to state $x$ via channel $\nu$. The transition rates $ k_{e}= k_{e}(\bm{f})$ along each edge $e$ completely defines the dynamics. Using the rates $k_{\exy}$, one can define: the total transition rate from $y$ to $x$ denoted $ k_{xy} \doteq \sum_{\nu}  k_{\exy}$, the escape rate from state $y$ via channel $\nu$ denoted $ \lambda_{(y,\nu)} \doteq \sum_{x \neq y}  k_{\exy}$, or the total escape rate from state $y$ denoted $ \lambda_{y}\doteq \sum_{\nu,x \neq y}  k_{\exy} $. All these rates are functions of the thermodynamic forces, for instance $\lambda_{y} = \lambda_{y}(\bm{f})$, even though we may omit it in the notation.  The rate matrix $\bm{ k} = \bm{ k}(\bm{f}) $ writes
\begin{equation}
	\bm{ k} \doteq \sum_{\nu} \left( \sum_{x,y\neq x} k_{(xy,\nu)} | x \rangle \langle y |  -  \sum_x \lambda_{(x,\nu)} | x \rangle \langle x | \right) \doteq  \sum_{\nu} \bm{k}_{\nu} 
\end{equation}
using bracket notation for the basis of the vector space of system states. We call $\bm{k}$ the rate matrix of the \emph{reference} dynamics and $\bm{k}_{\nu}$ the rate matrix for channel $\nu$. The state probability vector $\bm{p} = \sum_{x} p_{x} | x \rangle $ evolves according to the Markov master equation $\partial_{t} \bm{p} = \bm{k} \cdot \bm{p}$. If $\bm{f} \neq \bm{0}$ the system will eventually relax to a non-equilibrium stationary state. The rate matrix $\bm{k}$ can be used to generate trajectories of a stochastic jump process. A trajectory $[z]$ of duration $t$ gives the ordered list of visited states, the time interval between jumps, and the channel of each jump (if the transition can be done via different channels). From these trajectories, we define the time-averaged current $j_{e}[z]$ that counts the number of transitions along edge $e$ per unit time during the trajectory $[z]$. Many $[z]$ realizations of the stochastic process generate an ensemble of trajectory $\{ [z] \}$, from which an ensemble of random edge currents $\{ j_{e}[z] \}$ is produced. With a slight abuse of notation, we denote by $j_{\X}[z]$ the current of physical quantities $\X$ exchanged with the environment during a trajectory $[z]$. If $\phi_{\X,(xy,\nu)}$ denotes the amount of physical quantity $\X$ exchanged with reservoir $\nu$ during transition from $y$ to $x$, then the physical current along a trajectory writes
\begin{equation}
	j_{\X}[z]  = \sum_{e}  j_{e}[z]  \phi_{\X,e}. \label{physcurrent}
\end{equation}
We remark that since $j_{e}[z] = -j_{-e}[z]$ and $\phi_{\X,e} = -\phi_{\X,-e}$, an arbitrary orientation of edges used consistently leads to the same physical current.

We call fundamental currents those of the physical currents whose sole knowledge of the average (together with their conjugated thermodynamic forces $f_{\X}$) is sufficient to compute the total entropy production rate in the stationary state \cite{Polettini2016_vol94,Rao2018vol149}
\begin{equation}
	\l \sigma[z] \r_{\bm{k}} = \sum_{\X} b_{\X} \l j_{\X}[z] \r_{\bm{k}},
\end{equation}
where $\l \dots \r_{\bm{k}}$ denote the mean value on $\{ [z] \}$ generated by $\bm{k}$. Here, $\sigma[z]$ is the logarithm of the path probabilities of $[z]$ divided by the path probabilities of the time reverse trajectory, all divided by the trajectory duration. From now on, the subscript $\X$ labels only fundamental currents and their conjugated thermodynamic forces.

The moment generating function for fundamental currents observed on a time $t$ and conditioned on the final state, namely the vector of component
\begin{equation}
g_{x}(\bm{\alpha},\bm{f},t) \doteq \l e^{t \bm{\alpha} \cdot \bm{j}[z] } \delta(x-z(t)) \r_{\bm{k}(\bm{f})},
\end{equation} 
evolves according to \cite{Book_Sekimoto2010}
\begin{equation}
	\partial_{t} \bm{g} = \bm{\kappa} \bm{g},
\end{equation}
where we have introduced the tilted generator 
\begin{equation}
	\bm{\kappa}(\bm{\alpha},\bm{f}) \doteq \sum_{\nu,x,y\neq x} k_{(xy,\nu)}(\bm{f})  e^{\sum_{\X} \alpha_{\X} \hat \phi_{\X,(xy,\nu)}}| x \rangle \langle y |  -  \sum_x \lambda_x (\bm{f}) | x \rangle \langle x |,
	\label{TiltedOperator}
\end{equation}
where $\alpha_{\X}$ is the Laplace conjugated variable to physical current $j_{\X}$. 
Notice that, we have used a freedom for counting fundamental currents that is explained in Ref.~\cite{Andrieux2007_vol127,Altaner2015_vol92}, {hence not applying straightforwardly Donsker-Varadhan  techniques used to compute large-deviation of currents.}
The cycle exchange matrix $\hat {\bm \phi}$ gives the exchange of each physical quantity with the environment, i.e. its component $\hat \phi_{\X,\mathrm{c}i}$ is equal to the amount of physical quantity labeled by $\X$ that is exchanged during the fundamental cycle $\mathrm{c}i$. This matrix follows from the edge exchange matrix $\bm \phi$ of components $\phi_{\X,e}$ appearing in Eq.~\ref{physcurrent} and the matrix of fundamental cycles $\bm{C}$ as 
\begin{equation}
	\sum_e \phi_{\X,e} C_{e,\mathrm{c}i} = \hat \phi_{\X,\mathrm{c}i}.
\end{equation}
The column of $\bm{C}$ are the fundamental cycles of the system's graph \cite{Schnakenberg1976_vol48}. A cycle is a vector in the space of (arbitrarily) oriented edges. Entries of a cycle vector are $\pm 1$ (or $0$) if the edge belongs to the cycle (or not) with a sign chosen according to the edge orientation; the list of edges of a cycle shall form a loop, i.e. without ending states. Fundamental cycles represent a basis of the cycle vector space. In graph theory, one can build this basis starting from a spanning tree on the graph, i.e. a list of edges that connect all vertices of the graphs without creating a loop, and adding one at a time the remaining edges (called chords) to create fundamental cycles. Doing so, each chords is uniquely associated to a fundamental cycle \cite{Schnakenberg1976_vol48}.

The highest eigenvalue of the tilted matrix $\bm{\kappa}(\bm{\alpha},\bm{f})$ yields $\Gamma_{\bm{\kappa}}(\bm{\alpha},\bm{f})$ the Scaled Cumulant Generating Function (SCGF) of fundamental currents. This SCGF is connected to the moment generating functions conditioned on final state by
\begin{equation}
	\Gamma_{\bm{\kappa}}(\bm{\alpha},\bm{f}) \underset{t \to \infty}{=} \frac{1}{t} \ln \sum_{x} g_{x}(\bm{\alpha},\bm{f},t).
\end{equation}

In the next section, we define the so-called root dynamics whose tilted generator $\bm{\mathcal{R}}$ is related by a similarity transform and a translation on the Laplace variable to $\bm{\kappa}$. Hence the two tilted generators have identical spectrum. Equivalence classes for non-equilibrium processes are defined next since many dynamics share the same root dynamics.

\section{Similarity of root and reference dynamics}
\label{similarity}

We call \emph{root matrix} $\bm{r}=\bm{r}(\bm{f})$ the generator of the root dynamics defined as the sum of the geometric mean between the off-diagonal elements of $\bm{k}_{\nu}$ and $\bm{k}^{T}_{\nu}$, i.e.
\begin{equation}
	\bm{r} (\bm{f}) \doteq  \sum_{\nu,x,y\neq x } \sqrt{k_{(xy,\nu)} (\bm{f}) k_{(yx,\nu)}(\bm{f})}  | x \rangle \langle y |  -  \sum_x \lambda_x(\bm{f}) | x \rangle \langle x |, 
\end{equation}
while the diagonal elements are not modified. By definition, the root matrix is symmetric, but it is \emph{not} a Markov matrix. Based on the normalized left eigenvector of $\bm{r}$ associated to the highest eigenvalue, one can obtain a Markov matrix as the Doob transform of $\bm{r}$. Such a transformation is essentially a similarity transform of the rate matrix combined with a constant translation on the diagonal \cite{Chetrite2015_vol16,Jack2010_vol184}. The Doob transform of $\bm r$ will satisfy a detailed balance equation, and therefore has a stationary state which is an equilibrium state. 

The tilted root matrix writes
\begin{equation}
	\bm{\mathcal{R}}(\bm{a},\bm{f}) \doteq \sum_{\nu , x,y\neq x} \sqrt{k_{(xy,\nu)}(\bm{f}) k_{(yx,\nu)}(\bm{f}) }  e^{\sum_{\X} a_{\X} \hat \phi_{\X,(xy,\nu)}}| x \rangle \langle y |  -  \sum_x \lambda_x (\bm{f}) | x \rangle \langle x |, \label{TiltedRootMatrix}
\end{equation}
and we denote with $\Gamma_{\bm{\mathcal{R}}}(\bm{a},\bm{f})$ its highest eigenvalue.
To relate $\bm{\kappa}$ and $\bm{\mathcal{R}}$ by similarity transform, we introduce the drift potential $u=u(\bm{f})$ as
\begin{equation}
	\exp\left[\Delta u_e(\bm{f}) \right] \doteq \sqrt{\frac{k_{e}(\bm{f})}{k_{-e}(\bm{f})}} \text{ for any edge } e \text{ that is \emph{not} a chord}, \label{defdrifpot}
\end{equation}
where $\Delta u_{(xy,\nu)} = u_{x} - u_{y}$. This defines $\bm{u}$ up to an additive constant that we set by choosing a unit norm for $e^{\bm{u}}$. More generally, the drift potential satisfies
\begin{equation}
	\exp\left[\Delta u_e(\bm{f}) \right] = \sqrt{\frac{k_{e}(\bm{f})}{k_{-e}(\bm{f})}} \exp \left(-\sum_{\X} f_{\X} \hat \phi_{\X,e} \right) \text{ for any edge } e, \label{DriftPotProp}
\end{equation}
with a slight abuse of notation to extend the exchange matrix for cycles $\bm{\hat \phi}$ to the full edge space as follows
\begin{equation}
	\hat \phi_{\X,e} \doteq \left\{ 
\begin{array} {l}
    \hat \phi_{\X,\mathrm{c}i} \;\;\; \text{ if } e \text{ is the chord of cycle } \mathrm{c}i,  \\
    0 \qquad \;\; \text{ otherwise.}
\end{array} \right. \label{extended-exchange-matrix}
\end{equation}
Indeed, from the generalized detailed balance \cite{Schnakenberg1976_vol48}, half cycle affinities $\hat f_{\mathrm{c}i}$ are defined by
\begin{equation}
	\hat f_{\mathrm{c}i} \doteq \sum_{\X} f_{\X} \hat \phi_{\X,\mathrm{c}i} = \ln \left( \prod_{e \in \mathrm{c}i} \sqrt{\frac{k_e}{k_{-e}}} \right) , \label{cycleforces}
\end{equation}    
where the product is on all edges belonging to cycle $\mathrm{c}i$. Using the definition of drift potential in Eq.~(\ref{defdrifpot}) and specifying out the chord $e'$ in the product on the edges of cycle $\mathrm{c}i$, we obtain
\begin{equation}
	\sum_{\X} f_{\X} \hat \phi_{\X,e'} = \ln \left( \sqrt{\frac{k_{e'}}{k_{-e'}}} \prod_{e \in \mathrm{c}i \setminus e'} \sqrt{\frac{k_e}{k_{-e}}} \right) = \ln \sqrt{\frac{k_{e'}}{k_{-e'}}}  - \Delta u_{e'} ,
\end{equation}
that yields Eq.~(\ref{DriftPotProp}). From this latter equation and the definition of the tilted root matrix of Eq.~(\ref{TiltedRootMatrix}), one finds
\begin{equation}
	\bm{\kappa}(\bm{\alpha},\bm{f}) = \mathrm{D}  \left( e^{\bm{u}(\bm{f})} \right) \cdot \bm{\mathcal{R}}(\bm{f}+\bm{\alpha},\bm{f}) \cdot \mathrm{D} \left( e^{-\bm{u}(\bm{f})} \right), \label{SimKappaR}
\end{equation}
where $\mathrm{D}(\bm v)$ transforms vector $\bm v$ into a diagonal matrix. We notice that $e^{\bm u(\bm{f})}$ is the left eigenvector of $\bm{\mathcal{R}}(\bm{f},\bm{f})$ associated to null eigenvalue, since the  line vector with all entries equal to one is by definition the left eigenvector of
\begin{equation}
	\bm{k}(\bm{f}) = \bm{\kappa}(0,\bm{f}) = \mathrm{D}  \left( e^{\bm{u}(\bm{f})} \right) \cdot \bm{\mathcal{R}}(\bm{f},\bm{f}) \cdot \mathrm{D} \left( e^{-\bm{u}(\bm{f})} \right),
\end{equation}
for the null eigenvalue. Therefore, $\bm{k}(\bm{f})$ is the Doob transform of $\bm{\mathcal{R}}(\bm{f},\bm{f})$ based on vector $e^{\bm{u}(\bm f)}$. 
From Eq.~(\ref{SimKappaR}), we find that the spectra of the reference and root tilted matrices satisfy
\begin{equation}
	\Gamma_{\bm \kappa}(\bm{\alpha},\bm{f}) = \Gamma_{\bm{\mathcal{R}}}(\bm{f}+\bm{\alpha},\bm{f}). \label{SimilarSpectrum}
\end{equation}
With words, we interpret Eqs.~(\ref{SimKappaR}) and (\ref{SimilarSpectrum}) as follows: the counting field $\bm{a} = \bm{f} + \bm{\alpha}$ appearing in Eq.~(\ref{TiltedRootMatrix}) includes a tilting force $\bm{f}$ that transforms the equilibrium dynamics based on $\bm{r}$ into the driven dynamics with generator $\bm{\mathcal{R}}(\bm{f},\bm{f})$ (similar to $\bm{k}$ the generator of the reference dynamics) and a counting field $\bm{\alpha}$ that explores the current fluctuation of the reference dynamics.

We remark that the exchange $e \rightarrow -e$ and $\bm a \rightarrow -\bm a$ is a symmetry of Eq.~\ref{TiltedRootMatrix} and hence $\bm{\mathcal{R}}(\bm a, \bm f) = \bm{\mathcal{R}}(-\bm a , \bm f)$ which encodes the symmetry of current fluctuations at equilibrium. Using $\bm a = \bm f +\bm\alpha$ this translates into $\bm{\mathcal{R}}(\bm f + \bm \alpha, \bm f) = \bm{\mathcal{R}}(\bm f-2\bm f -\bm\alpha , \bm f) $ and Eq.~\ref{SimKappaR} produces for the tilted generator of the reference dynamics 
\begin{equation}
	\bm\kappa(\bm\alpha,\bm f) = \bm\kappa(-2\bm f-\bm\alpha,\bm f)
\end{equation}
that would follow directly from Eq.~\ref{TiltedOperator} and \ref{cycleforces} as well. Therefore the spectrum of these two matrices are identical and more specifically one gets the asymptotic fluctuation relation for fundamental currents
\begin{equation}
	\Gamma_{\bm \kappa}(\bm \alpha,\bm f) = \Gamma_{\bm \kappa}(-2\bm f - \bm \alpha,\bm f).
\end{equation}
{One concludes that the fluctuation relation brings to all elements of a dynamical equivalence class the symmetry of current fluctuations of the equilibrium dynamics in the same dynamical equivalence class, namely the root dynamics. Hence, an alternative and less convenient way of defining the root dynamics is to constrain the reference dynamics on having null entropy production.}

\section{{Dynamical} equivalence classes }
\label{equivalence}

Since the statistics of fundamental currents is fully determined by the spectrum of the tilted root matrix, two jump processes with the same components of the root matrix on all edges and on the diagonal share the same fluctuations of fundamental currents {in the stationary state. This implies that proportionality on each edge and on the diagonal leads to proportional SCGF as well}. Accordingly, the following equivalence relation
\begin{equation}
	\bm k \equiv \bm k' \text{ if } \forall e, \sqrt{k_{e}k_{-e}} = \gamma \sqrt{k'_{e}k'_{-e}} \text{ and if } \forall y, \sum_{x \neq y,\nu} k_{xy,\nu} = \gamma \sum_{x\neq y,\nu} k_{xy,\nu}'.
\end{equation} 
defines a {dynamical} equivalence class between two jump processes. We emphasize that a couple of states may be connected by several channels and that the component of the root matrices must be proportional for each channel separately. The above equivalence class extends to jump processes on multigraphs and in continuous time the equivalence class introduced by Andrieux in \cite{Andrieux2012_vola} for Markov chains (discrete time) on simple graphs. As for Markov chains, the spectra of the tilted rate matrices associated to $\bm k$ and $\bm k'$ in the same equivalence class are proportional. {We remark that proportionality of the spectra is stronger than of the SCGF as it is associated to path ensemble equivalence (at finite time) upon a suitable choice of boundary conditions.}

The difference between two matrices $\bm k$ and $\bm k'$ in the same equivalence class can arise from the different ways of breaking detailed balance. Generalized detailed balance does so by defining differently the skew symmetric part of these matrices. For instance, forces may appear on every edges
\begin{equation}
	\frac{k_{e}(\bm{f})}{k_{-e}(\bm{f})} {\propto} e^{\sum_{\X}f_{\X} \phi_{\X,e}},
\end{equation}
or on chords only 
\begin{equation}
	\frac{k'_{e}(\bm{f})}{k'_{-e}(\bm{f})} {\propto} e^{\sum_{\X}f_{\X} \hat \phi_{\X,e}},
\end{equation}
which means that the forward and backward rates are the same for any other edge {(up to a potential contribution)}. Notices that the cycle forces $\bm {\hat f}$ defined in Eq.~\ref{cycleforces} are the same defining them with $\bm k$ or with $\bm k'$. Since these two rate matrices are in the same equivalence class, they are both related by similarity transform to $\bm{ \mathcal{ R}}(\bm f, \bm f)$, but through two different drift potentials, respectively the one defined in Eq.~\ref{defdrifpot} and the one defined by the same equation with $\bm k$ and $\bm u$ replaced by $\bm k'$ and $\bm u'$.

Interestingly, for equilibrium stationary state, the symmetric part of the rate matrix has no influence on thermodynamic equilibrium and just the detailed balance equation for the skew symmetric part matters. On the opposite, the nonequilibrium stationnary state when characterized by the fundamental currents fluctuations is only determined by the symmetric part of the rate matrix (defining the equivalence class), given that the skew symmetric part has the expected fundamental forces. In this sense, when dealing with non-equilibrium stationary jump processes, there is a dynamical freedom in the way of imposing the fundamental forces that has no consequences on the fluctuations of fundamental currents. This dynamical freedom is in fact a gauge freedom since in the framework of path probability the Doob transform relating two elements in an equivalence class is in fact a gauge change~\cite{Chabane2021_vol}.

Nonequilibrium systems change of equivalence class when modifying the thermodynamic forces since these forces modify at least the diagonal part of the root matrix which is related to the system activity \cite{Maes2020vol850}. In most cases, forces also appear in the off-diagonal part of the root matrix. For instance in photoelectric or thermoelectric devices, transition rates often follow from from quantum perturbation analysis producing rates that depend on the Fermi (Bose) statistics of the electrons (bosons or phonons) reservoirs. In these cases, the thermodynamics forces such as the temperature and electrochemical potential differences appear in the symmetric part of the rate matrix and hence in the root matrix. It is a challenge to develop a nonequilibrium thermodynamics of such non trivial systems by including activity into a coherent thermodynamics structure.

\section{Illustration on a discrete model of molecular motor}

\begin{figure}[b]
  \centering
  \includegraphics[scale=0.6]{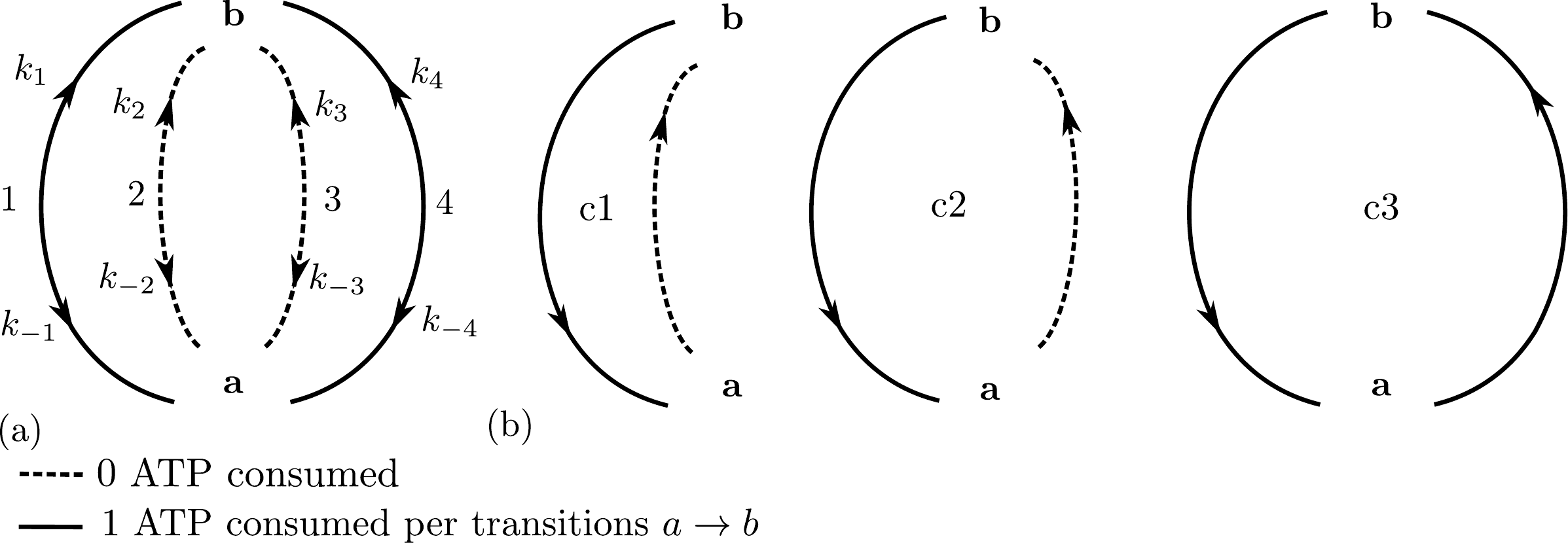}
  \caption{(a) Sketch of the effective two-state system with four edges. Edge orientation is head toward $b$. (b) Set of fundamental cycles with their orientations given by the orientation of their corresponding chords. The $i$th column of matrix $\bm C$ in Eq.~\ref{cyclematrix} is for cycle $\mathrm{c}i$. \label{fig:sketch-molecular-motor}}
\end{figure}

For clarity, we illustrate our main results on a discrete model of molecular motor described in Ref.~\cite{Lau2007_vol99,Lacoste2008_vol78, Vroylandt2018vol2018}. This isothermal molecular motor has two internal states $a$ and $b$ of energy $0$ and $\epsilon$ respectively. The two internal states are connected by four different microreversible transitions: the motor can move one step to the left or to the right with or without consuming Adenosine TriPhosphate (ATP) molecules. 
We associate as shown in figure \ref{fig:sketch-molecular-motor} a positive edge number $e=1,2,3 $ and $4$ to each transition from $a$ to $b$ and write $k_{e}$ the corresponding rate. Edge $1$ (resp. $2$) corresponds to the motor leaving state $a$ by moving to the \emph{left} while consuming $1$ (resp. $0$) ATP molecule. Edge $3$ (resp. $4$) corresponds to the motor leaving state $a$ by moving to the \emph{right} while consuming $0$ (resp. $1$) ATP molecule. The dimensionless fundamental forces $f_{\X}$ in this model are $f_{\I}=\Delta \mu /(2k_{B}T)$ that is half the chemical potential difference $\Delta \mu$ of the hydrolysis reaction of ATP and $f_{\II} = fd/(2k_{B}T)$ that is half the mechanical work $fd$ that the motor performs against force $f$ to make a step of length $d$. The fundamental currents are the ATP consumption rate $j_{\I}$ and the number of step per unit time $j_{\II}$. Hence, the total entropy production rate writes $\sigma = 2j_{\I}f_{\I}+2j_{\II}f_{\II}$. For simplicity, we write energies in thermal unit $k_{B}T =1$ where $T$ is the temperature of the motor environment. We take the Boltzmann constant $k_{B}=1$: the entropy production rates is homogeneous to an inverse time like fundamental currents. 

We completely specify the dynamics using the following transition rates
\begin{equation}
  \begin{array}{ll}
 k_{-1} = \tilde \alpha e^{2\theta^{+}_bf_{\II}}, & k_{-2} =  \tilde \omega\,e^{2\theta^{+}_b f_{\II}},\\
 k_{1}  =  \tilde \alpha e^{ -\epsilon + 2f_{\I} - 2\theta^{-}_a f_{\II}}, & k_{2} = \tilde \omega\,e^{-\epsilon- 2\theta^{-}_af_{\II}},\\
 k_{-4} = \alpha\, e^{-2\theta^{-}_bf_{\II}}, & k_{-3} =  \omega\,e^{-2\theta^{-}_bf_{\II}},  \\
 k_{4}  =  \alpha\, e^{ -\epsilon + 2f_{\I} + 2\theta^{+}_a f_{\II} }, & k_{3}= \omega\,e^{-\epsilon + 2\theta^{+}_a f_{\II}} \, ,
\end{array}
\label{eq:ratesMolecularMotor}
\end{equation}
where $\alpha$, $\tilde \alpha$, $\omega$ and $\tilde \omega$ are time scales for the various transitions, and $\theta_{x}^{\pm}$ are the load  distribution factors that encodes the left/right asymmetry of the motor (inherited from the modeling in a continuous state space). These factors are arbitrary except for the constraint $\theta^{+}_a+\theta^{-}_b + \theta^{-}_a+\theta^{+}_b =2$. {This constraint ensures the thermodynamic consistency of the model: the entropy production computed using edge currents and forces is equal to the entropy production computed using fundamental current and forces \cite{Lacoste2008_vol78,Polettini2016_vol94}.} The root matrix writes 
\begin{equation}
  \bm r (\bm f) = 
  \left( \sum_{e=1}^{4} r_{e} \right) | a \rangle \langle b | + \left( \sum_{e=1}^{4}  r_{e} \right) | b \rangle \langle a | - \left(\sum_{e=1}^{4} k _{e}\right) | a \rangle \langle a | - \left(\sum_{e=1}^{4} k _{-e}\right) | b \rangle \langle b |
\label{eq:roorratesMolecularMotor}
\end{equation}
in term of the symmetrized rates $r_{e} = \sqrt{k_{e}k_{-e}}$ that are explicitely:
\begin{equation}
  \begin{array}{ll}
  r_{1} = r_{-1} = \tilde \alpha e^{ -\epsilon/2 + f_{\I} + (\theta^{+}_b-\theta^{-}_a) f_{\II}}, 	
& r_{2} = r_{-2} = \tilde \omega\,e^{-\epsilon/2+ (\theta^{+}_b-\theta^{-}_a)f_{\II}},\\
  r_{4} = r_{-4} = \alpha\, e^{ -\epsilon/2 + f_{\I} + (\theta^{+}_a-\theta^{-}_b) f_{\II} }, 	
& r_{3} = r_{-3} = \omega\,e^{-\epsilon/2 + (\theta^{+}_a-\theta^{-}_b) f_{\II}} \,.
  \end{array}
\label{eq:RootratesMolecularMotor}
\end{equation}

We define now the cycle and exchange matrices in order to build the tilted root matrix $\bm{\mathcal{R}}(\bm \alpha,\bm f)$. There are four different spanning trees for the graph of this model of molecular motor since its spanning trees correspond here to just one edge. We chose as spanning tree the edge $1$. Adding an edge to this spanning tree creates one cycle. We use the following cycle matrix 
\begin{equation}
    \bm{C} = \left( \begin{array}{ccc}
    -1 & -1 &-1 \\
    1& 0 &0 \\
    0 &1 &0 \\
    0 & 0 &1
    \end{array} \right), \label{cyclematrix}
\end{equation}
for cycles $\mathrm{c}1$, $\mathrm{c}2$ and $\mathrm{c}3$ defined in figure \ref{fig:sketch-molecular-motor} and corresponding to the first, second and third column of $\bm C$. Notice that chord orientation determines the orientation of its associated fundamental cycle. Since, we assume positive edge orientation for transitions from $a$ to $b$, chords are oriented in the same way from $a$ to $b$ and hence cycle $\mathrm{c}1$ is made of edges $2$ and $-1$, cycle $\mathrm{c}2$ is made of edges $3$ and $-1$, etc. To convert probability currents in edge space to physical currents, we need to define the edge exchange matrix whose line $\I$ (resp. $\II$) provides the number of consumed ATP molecule (resp. the number of steps to the right done by the motor) when the $e$th transition occurs, with $e$ being the column index,
\begin{equation}
  \label{eq:MMedegcontribution}
 \bm{ \phi} =   \left( \begin{array}{cccc}
     1 & 0 & 0 & 1 \\
    -1 &-1 & 1 & 1 \\
  \end{array} \right).
\end{equation}
Then, summing the exchanges along every edges of each fundamental cycle gives the cycle exchange matrix 
\begin{equation}
\label{matrix_phi}
  \bm{\hat \phi}  = \bm{ \phi} \cdot \bm{C} =
  \left( \begin{array}{ccc}
           -1 & -1 & 0\\
            0 &  2 & 2 \\
         \end{array} \right) \text{ in cycle space.}
\end{equation}
To define this matrix of components $\hat \phi_{\X,\mathrm{c}i}$ into the space of edges instead of fundamental cycles as done in Eq.~\ref{extended-exchange-matrix}, we extend it on the left with one more column filled with zeros since edge $1$ is not a chord, while edges $2,3$ and $4$ are respectively the chords of cycles $\mathrm{c}1$, $\mathrm{c}2$ and $\mathrm{c}3$: 
\begin{equation}
  \bm{\hat \phi} =
  \left( \begin{array}{cccc}
          0 &-1 & -1 & 0\\
          0 & 0 &  2 & 2 \\
         \end{array} \right) \text{ in edge space.}
\end{equation}
The drift potential is defined thanks to edge $1$ (the unique edge of our spanning tree) and for all edges $e=1,2,3,4$ (hence going from $a$ to $b$) we have
\begin{equation}
	\Delta u_{e} = u_{b} - u_{a} = \ln \sqrt{ \frac{k_{1}}{k_{-1}} } = -\epsilon/2 + f_{\I} -( \theta^{-}_a + \theta^{+}_b)f_{\II}.\label{drift-motor}
\end{equation}
We emphasize that the rates of Eq.\ref{eq:ratesMolecularMotor} satisfy Eq.\ref{DriftPotProp} since
\begin{eqnarray}
	\ln \sqrt{ \frac{k_{2}}{k_{-2}} } &=& -\epsilon/2 - (\theta^{-}_a+\theta^{+}_b ) f_{\II} =  u_{b} - u_{a} - f_{\I},\\
	\ln \sqrt{ \frac{k_{3}}{k_{-3}} } &=& 
	 -\epsilon/2 + (2-\theta^{-}_a-\theta^{+}_b ) f_{\II} =  u_{b} - u_{a}-f_{\I} + 2f_{\II}, \\
	\ln \sqrt{ \frac{k_{4}}{k_{-4}} } &=& 
	-\epsilon/2 + f_{I} + (2-\theta^{-}_a-\theta^{+}_b ) f_{\II} =  u_{b} - u_{a} + 2f_{\II},
	\label{check-drift}
\end{eqnarray}
where we used $\theta^{+}_a+\theta^{-}_b + \theta^{-}_a+\theta^{+}_b\ =2~$ for the last two lines.

\begin{figure}
  \centering
  \includegraphics[scale=0.6]{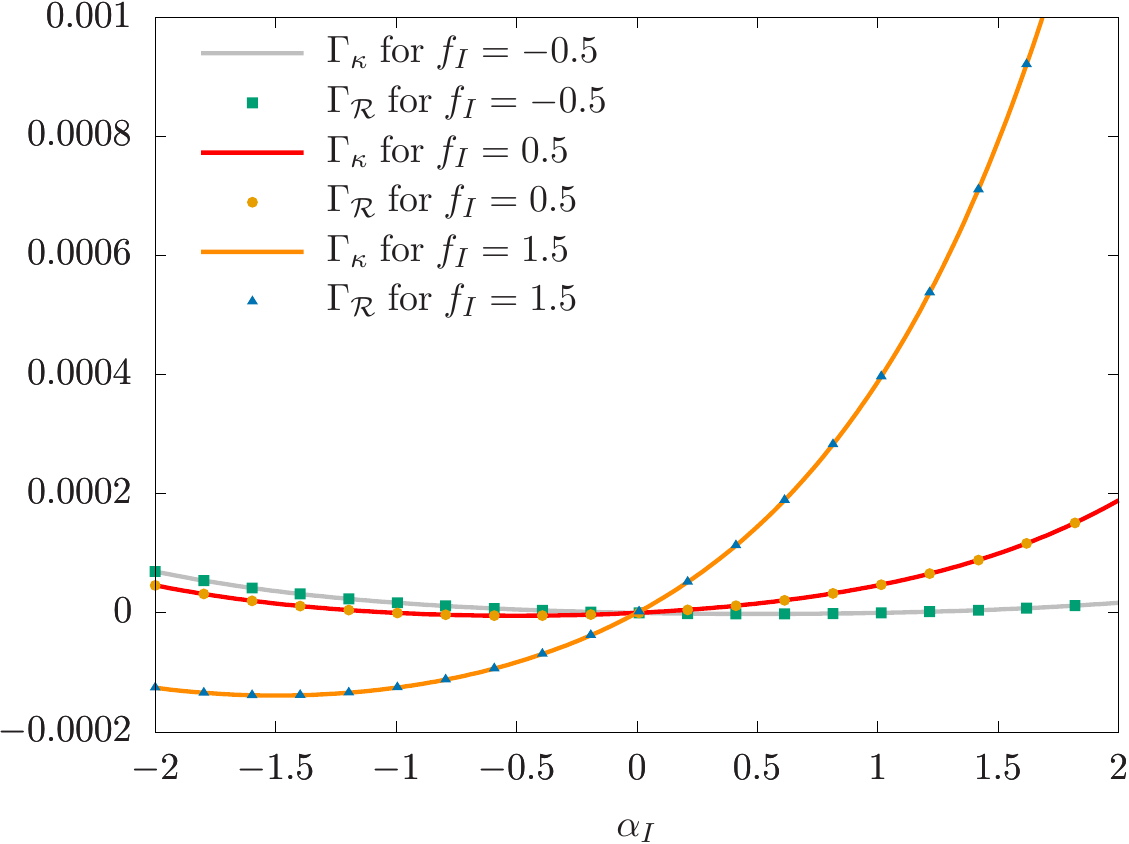}
  \includegraphics[scale=0.6]{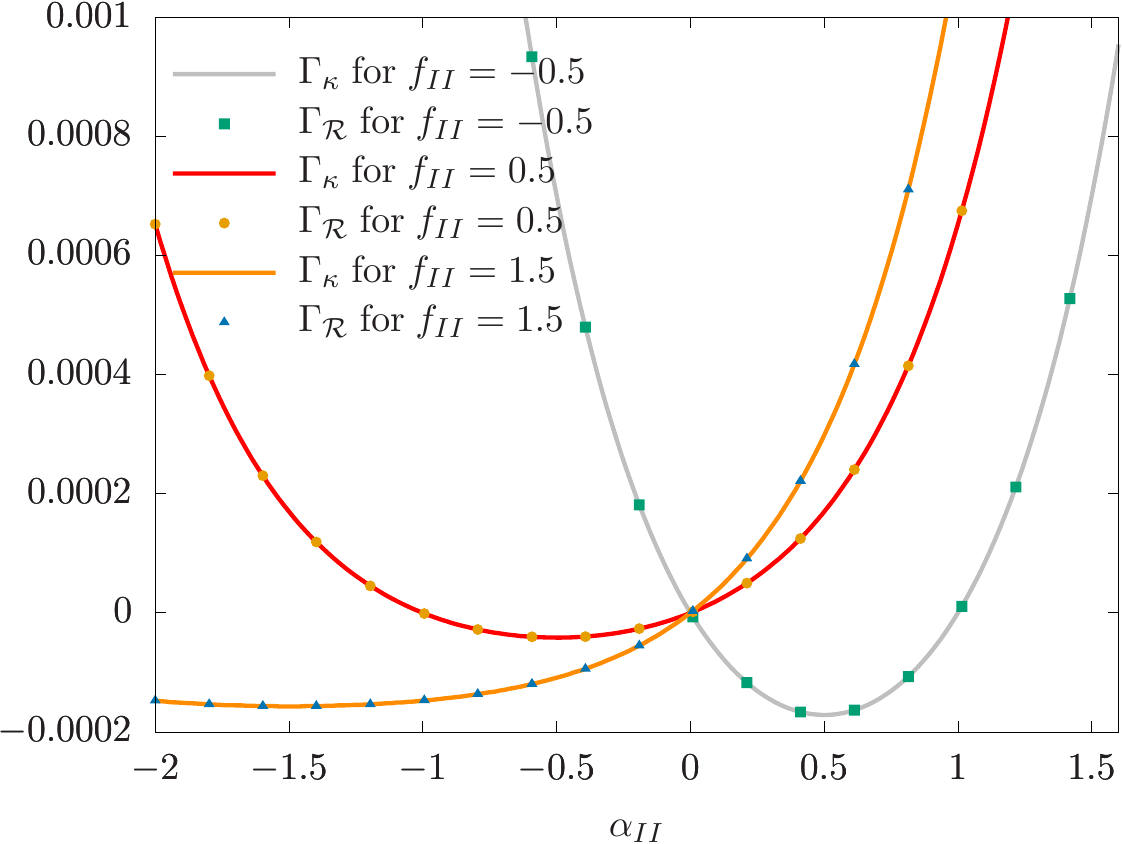}
  \caption{Comparison of the spectra of the tilted root matrix $\mathcal{R}$ with the one of the tilted rate matrix $\kappa$. (Left) Spectra $\Gamma_{\kappa}(\alpha_\textsc{i},\alpha_\textsc{ii} = 0,f_\textsc{i},f_\textsc{ii}=0)$ and $\Gamma_{\mathcal{R}}(f_\textsc{i}+\alpha_\textsc{i},\alpha_\textsc{ii} = 0,f_\textsc{i},f_\textsc{ii}=0)$ as a function of $\alpha_\textsc{i}$ for $f_\textsc{i}=-0.5$, $0.5$ and $1.5$. (Right) Spectra $\Gamma_{\kappa}(\alpha_\textsc{i} = 0,\alpha_\textsc{ii},f_\textsc{i}=0,f_\textsc{ii})$ and $\Gamma_{\mathcal{R}}(\alpha_\textsc{i} = 0, f_\textsc{ii}+\alpha_\textsc{ii},f_\textsc{i}=0,f_\textsc{ii})$ as a function of $\alpha_\textsc{ii}$ for $f_\textsc{ii}=-0.5$, $0.5$ and $1.5$.   \label{fig:spectra} {For both figures, we took $\alpha=0.57$, $\tilde \alpha=1.3\,10^{-6}$, $\omega =3.5$, $\tilde \omega=108.15$, $\epsilon =10.81$, $\theta_{a}^{+}= 0.25$, $\theta_{a}^{-}= 0.75$, $\theta_{b}^{+}=0.75 $, $\theta_{b}^{-}= 0.25$.}}
\end{figure}

{First, we illustrate Eq.~\ref{SimilarSpectrum} relating the spectrum of the tilted rate matrix to the spectrum of the tilted root matrix.} With the cycle exchange matrix written in the edge space, the tilted root matrix reads
\begin{eqnarray}
  \bm{\mathcal{R}}(\bm \alpha,\bm f) &=& 
  \left( \sum_{e=1}^{4}  r_{e} e^{\sum_{\X} \alpha_{\X}\hat \phi_{\X,e}} \right) | b \rangle \langle a | + \left( \sum_{e=1}^{4}  r_{e}e^{-\sum_{\X} \alpha_{\X}\hat \phi_{\X,e}} \right) | a \rangle \langle b | \nonumber\\ && - \left(\sum_{e=1}^{4} k _{e}\right) | a \rangle \langle a | - \left(\sum_{e=1}^{4} k _{-e}\right) | b \rangle \langle b |
\end{eqnarray}
whose highest eigenvalue $\Gamma_{\bm{\mathcal{R}}} = \Gamma_{\bm{\mathcal{R}}}(\bm \alpha,\bm f)$ writes in term of its trace and determinant as
\begin{equation}
	\Gamma_{\bm{\mathcal{R}}} = \frac{1}{2}\left( \mathrm{Tr}\, \bm{\mathcal{R}} +\sqrt{[\mathrm{Tr}\, \bm{\mathcal{R}}]^{2}-4\mathrm{Det}\, \bm{\mathcal{R}}} \right).
\end{equation}
Similarly, tilting the rate matrix $\bm k$ gives
\begin{eqnarray}
  \bm{\kappa}(\bm \alpha,\bm f) &=& 
  \left( \sum_{e=1}^{4}  k_{e} e^{\sum_{\X} \alpha_{\X}\hat \phi_{\X,e}} \right) | b \rangle \langle a | + \left( \sum_{e=1}^{4}  k_{-e}e^{-\sum_{\X} \alpha_{\X}\hat \phi_{\X,e}} \right) | a \rangle \langle b | \nonumber\\ && - \left(\sum_{e=1}^{4} k _{e}\right) | a \rangle \langle a | - \left(\sum_{e=1}^{4} k_{-e}\right) | b \rangle \langle b |
\end{eqnarray}
whose trace and determinant are by construction exactly the same as those of $\bm{\mathcal{R}}(\bm \alpha+\bm f,\bm f)$. This can be checked directly using Eqs.~\ref{drift-motor}--\ref{check-drift} illustrating Eq.~\ref{SimilarSpectrum} for the identity of the two spectra (when translating the counting field $\bm \alpha$ with the force $\bm f$). Alternatively, we illustrate this relation on figure \ref{fig:spectra}.

{Second, we illustrate the notion of dynamical equivalence class by finding two models of molecular motor having the same SCGF for currents $j_{\I}$ and $j_{\II}$, up to a multiplicative constant $\gamma$ relating the time scales of the two models. As we saw in section~\ref{equivalence}, this is possible if each root rate and escape rate of the two models are proportional with the same proportionality constant $\gamma$. For the molecular motor introduced above, a well chosen modification of the load factors $(\theta^{\pm}_{a},\theta^{\pm}_{b})$ and energy $\epsilon$ allows to change of dynamics while remaining in the same dynamical equivalence class.}

{The modification of the load factors must be done in agreement with the thermodynamic consistency $ \theta^{+}_a+\theta^{-}_b + \theta^{-}_a+\theta^{+}_b = 2$ and we chose to keep $ \theta_{b}^{+}-\theta_{a}^{-} $ and $ \theta_{a}^{+}-\theta_{b}^{-} $ constant. 
Introducing two parameters $\theta_{a}$ and $\theta_{b}$ used to move inside the dynamical equivalence class, the load factors can be set by
\begin{equation}
	\theta_{a}^{\pm}  = \frac{1\pm\theta_{a}}{2}, \qquad \theta_{b}^{\pm} = \frac{1\pm\theta_{b}}{2} \quad \Rightarrow \quad \theta_{b}^{+}-\theta_{a}^{-} = \theta_{a}^{+}-\theta_{b}^{-} = \frac{\theta_{a}+\theta_{b}}{2}.
\end{equation}
Then, the escape rates from state $a$ and $b$ read respectively
\begin{eqnarray}
	\sum_{e=1}^{4} k_{e} &=& e^{-\epsilon+(1+\theta_{a})f_{\II}}\left( \tilde \alpha e^{2(f_{\I}-f_{\II})} + \tilde \omega e^{-2f_{\II}} + \alpha e^{2f_{\I}} + \omega \right) \label{escapeA}, \\ 
	\sum_{e=1}^{4} k_{-e} &=& e^{(1+\theta_{b})f_{\II}}\left( \tilde \alpha +\tilde \omega + \alpha e^{-2f_{\II}} + \omega e^{-2f_{\II}}\right), \label{escapeB}
\end{eqnarray}
while any root rate behaves as
\begin{equation}
	r_{\pm e} \propto e^{-\epsilon/2 + f_{\II}(\theta_b+\theta_a)/2 }, \label{rootPROP}
\end{equation}
where the proportionality constant is independent of the load factors and of the energy. From Eqs.~(\ref{escapeA}--\ref{rootPROP}), we notice that the change
\begin{eqnarray}
	\theta_{a} &\rightarrow& \theta_{a} + \Delta \theta_{a} \label{modifthetaa} \\
	\theta_{b} &\rightarrow& \theta_{b} + \Delta \theta_{b} \\
	\epsilon &\rightarrow& \epsilon + f_{\II}(\Delta \theta_{a}-\Delta \theta_{b}) \label{modifepsilon}
\end{eqnarray}
produces an edgewise modification of the root dynamics by just a multiplicative factor $\gamma = e^{\Delta \theta_{b} f_{\II}}$. Therefore, modifying the dynamics according to Eqs~(\ref{modifthetaa}--\ref{modifepsilon}) is a non-trivial displacement within a dynamical equivalence class. This modification is not trivial in the sense that it does not just multiply the transition rates of Eq.~(\ref{eq:ratesMolecularMotor}) by a constant factor.}

\section*{Acknowledgment}

I thank Luca Peliti for organizing the workshop ``Random Talks on Stochastic and Non-Equilibrium Thermodynamics'' for which most of this work was prepared.

\appendix
\addcontentsline{toc}{section}{Appendices}

\section{Dynamics with same mean currents}
\label{appendix-MeanEquiv}

\emph{Symmetric freedom---} For sack of completeness, we review in this appendix former works on non-equilibrium equivalence class at the level of mean currents. As mentioned in the introduction, Zia and Schmittmann have shown that an \emph{additive} and symmetric modification of the jump probabilities does not modify the stationary state probability and the probability current \cite{Zia2007_vol}. To see it, let's define $p^\mathrm{st}_{y}$ the stationary probability of state $y$ such that $\sum_{y} k_{xy} p^\mathrm{st}_{y} = 0$. Let's define the symmetric matrix $\bm{h}$ with off-diagonal elements $h_{xy} $ (and diagonal elements $h_{yy} = -\sum_{x\neq y} h_{xy}$) chosen such that the new rate matrix $k'_{xy} = k_{xy} + h_{xy}/p^\mathrm{st}_{y}$ has positive off-diagonal elements. Then, the stationary state probability for matrices $\bm k$ and $\bm k'$ are the same since
\begin{eqnarray}
	\sum_{y} k'_{xy} p^\mathrm{st}_{y} &=& \sum_{y} \left(k_{xy} + h_{xy}/p^\mathrm{st}_{y} \right) p^\mathrm{st}_{y}, \\
	&=& \sum_{y} k_{xy} p^\mathrm{st}_{y} + \sum_{y} h_{xy}, \\
	&=& \sum_{y} h_{xy} = \sum_{x} h_{xy} = 0,
\end{eqnarray}
where we used the symmetry of $\bm h$ and the fact that the sums of its column elements are zero. Similarly, the probability currents associated to rate matrices $\bm k$ and $\bm k'$  are the same
\begin{eqnarray}
	 j^\mathrm{st'}_{xy} &\doteq& k'_{xy} p^\mathrm{st}_{y} - k'_{yx} p^\mathrm{st}_{x}, \\
	  &=& k_{xy} p^\mathrm{st}_{y} + h_{xy} - k_{yx} p^\mathrm{st}_{x} - h_{yx}, \\
	&=& k_{xy} p^\mathrm{st}_{y} - k_{yx} p^\mathrm{st}_{x} = j^\mathrm{st}_{xy},
\end{eqnarray}
again by symmetry of matrix $\bm h$. Therefore, if a nonequilibrium stationary state is characterized by the state probability $p^\mathrm{st}_{x}$ and all probability currents $j^\mathrm{st}_{xy}$, then the symmetric part of the rate matrix (without the diagonal) remains free (given that it is still a well defined rate matrix). Hence, there is just as much freedom in choosing rates to reach a given nonequilibrium stationary state as for systems to reach a thermal equilibrium state (since detailed balance condition associated to equilibrium state imposes no constraint either on the symmetric part of the rate matrix). Notice that the state probability and the probability currents enable to compute mean physical currents but no higher moments.

\emph{Waiting-times freedom---} When characterizing a non-equilibrium stationary state by the probability currents only, we find one additional dynamical degree of freedom for each state in comparison with the situation above. By construction, this freedom in the rate matrix corresponds to the arbitrariness of the stationary probability $\bm p^\mathrm{st}$. Indeed, introducing new transition rates $k_{xy}' = k_{xy} e^{\varphi_{y}}$, with $ \bm \varphi $ an arbitrary state vector, leads to the new stationary probability with components $p^\mathrm{st'}_{x} = p^\mathrm{st}_{x} e^{-\varphi_{x}}$ since 
\begin{equation} 
\sum_{y} k'_{xy} p^\mathrm{st'}_{y} = \sum_{y} \left( k_{xy} e^{\varphi_{y}} \right) \left( e^{-\varphi_{y}} p^\mathrm{st'}_{y}\right) =0.
\end{equation} 
As required, all probability currents are preserved $k'_{xy} p^\mathrm{st'}_{y} - k'_{yx} p^\mathrm{st'}_{x} =  k_{xy} p^\mathrm{st}_{y} - k_{yx} p^\mathrm{st}_{x}$. However, the escape rates become $\lambda'_{y} = \lambda_{y}e^{\varphi_{y}}$ and we may interpret the freedom on the stationary probability as a waiting-time freedom. This interpretation is confirmed by the fact that $\forall x, \;p^\mathrm{st}_{x} \propto n_{x} / \lambda_{x}$, where $n_{x}$ is the stationary probability of the embedded Markov chain (for very long trajectories, it is also the number of visit of state $x$ divided by the total number of jumps). From this proportionality relation, we find that the stationary probability of the embedded chain is also preserved when switching from $\bm k$ to $\bm k'$
\begin{equation}
	n_{x} = \frac{p^\mathrm{st}_{x}\lambda_{x}}{\sum_{y} p^\mathrm{st}_{y}\lambda_{y}} = \frac{p^\mathrm{st'}_{x}\lambda'_{x}}{\sum_{y} p^\mathrm{st'}_{y}\lambda'_{y}}.
\end{equation}
Hence, the modification of the stationary probability of the continuous time Markov jump process only comes from the freedom on the escape rates. Interestingly, Polettini has shown that more general transformations such as $k_{xy}' = k_{xy}v_{xy} e^{\varphi_{y}}$ will not lead to $p^\mathrm{st'}_{x} = p^\mathrm{st}_{x} e^{-\varphi_{x}}$ excepted if $v_{xy} $ is independent of $x$ and $y$ \cite{Polettini2012_vol97}. In this work, the modification of the stationary probability is referred to as a gauge freedom. However, the terminology ``gauge freedom'' should be used for the dynamical changes inside the equivalence classes introduced in the main text.

\section{Equivalence of edge-wise or cycle-wise tilting}
\label{edge-cycle-wise}

In this appendix, we provide another point of view on the edge and cycle freedom in the counting statistics of physical currents developed in Ref.~\cite{Wachtel2015_vol92}. In the stationary state, the statistics of fundamental currents is the same whether the physical quantity is counted for every transitions or for transitions associated to chords only, but with weights determined by the exchanged quantity during the corresponding fundamental cycle. In our framework, this freedom comes from the possibility of relating one tilting procedure to the other one by moving inside the equivalence class. 

To show this, we introduce many drift potentials $\bm u^{\beta}$ defined thanks to the many possibilities for choosing a spanning tree $T^{\beta}$ labeled by $\beta$:
\begin{equation}
	e^{\Delta u^{\beta}_{e}(\bm \alpha)} \doteq \sqrt{ \frac{ k_{e}(\bm \alpha)}{ k_{-e}(\bm \alpha)}} \exp \left(-\sum_{\X} \alpha_{\X} \hat \phi^{\beta}_{\X,e} \right) 
\end{equation}
with $\hat \phi^{\beta}_{\X,e} = \sum_{e'} \phi_{\X,e'} C^{\beta}_{e',\mathrm{c}i}$ if $e = \mathrm{c}i$ and $0$ otherwise, and where $\bm C^{\beta}$ is the matrix of fundamental cycles associated to the spanning tree $T^{\beta}$. Then, assuming $m^{\beta}$ are for now arbitrary real numbers, the off-diagonal components of the tilted matrix when counting on edges can be written 
\begin{eqnarray}
\fl	k_{e}(\bm f)e^{\sum_{X}\alpha_{\X} \phi_{\X,e}} &=& k_{e}(\bm f) e^{\sum_{X}\alpha_{\X} \phi_{\X,e}} \prod_{\beta}\left[ \sqrt{ \frac{ k_{e}(\bm \alpha)}{ k_{-e}(\bm \alpha)}}e^{-\Delta u^{\beta}_{e}(\bm \alpha) -\sum_{\X} \alpha_{\X} \hat \phi^{\beta}_{\X,e}} \right]^{m^{\beta}}, \\
	&=& \underbrace{ k_{e}(\bm f) \left[\frac{ k_{e}(\bm \alpha)}{ k_{-e}(\bm \alpha)}\right]^{m^{\beta}/2} e^{-\sum_{\beta} m^{\beta} \Delta u^{\beta}_{e}(\bm \alpha) }}_{\text{inside equivalence class of } \bm k(\bm f)} e^{\sum_{X} \alpha_{\X} \left[ \phi_{\X,e} - \sum_{\beta} m^{\beta} \hat \phi^{\beta}_{\X,e} \right] }. 
\end{eqnarray}
Hence, finding $m^{\beta}$ such that $  \sum_{\X} \alpha_{\X} \left[\phi_{\X,e} - \sum_{\beta }m^{\beta} \hat \phi^{\beta}_{\X,e}\right]  = \sum_{\X} \alpha_{\X} \hat \phi_{\X,e}$ would map the edge tilting into a chord tilting of another matrix that belongs to the equivalence class of $\bm k(\bm f)$. The problem of finding $m^{\beta}$ writes as a linear algebra problem
\begin{equation}
	\sum_{X} \alpha_{\X} \left[ \phi_{\X,e} -  \hat \phi_{\X,e}  \right] =   \sum_{\beta} m^{\beta} \left( \sum_{\X} \hat \phi^{\beta}_{\X,e} \right)
\end{equation}
where one needs to find the inverse of matrix $ \hat{\bm \Phi}$ with $(\beta,e)$ components equal to $\sum_{\X} \hat \phi^{\beta}_{\X,e}$. The index $\beta$ runs on the number of spanning trees. This number is $1$ for a tree graph (but switching from edge to cycle counting is meaningless in this case). It is equal to the number of edges on a cycle graph. It will be higher than the number of edges involved in all the cycles for an arbitrary connected graph. Then, there is enough free parameters $m^{\beta}$ to find an inverse for all edges that belongs to cycles of the graph. If matrix $\hat{\bm \Phi} $ is singular, then one can reduce the number of free parameter $m^{\beta}$ to find its inverse. In the end, using the freedom on the basis of fundamental cycles, on may bias a dynamics using the chords in many different ways that, together, can produce the desired bias on each edge.

Let's notice that edges belonging to tree parts of a graph do not contribute to current statistics at all: any physical quantity exchanged while entering into such a tree part will be exchanged back when leaving this tree part. Algebraically, the exponential tilting on edges that are on the tree part of the graph disappears upon symmetrization of the tilted rate matrix. They just contribute to the skew symmetric part of the rate matrix and just change the particular representative in the class that is being tilted.

\vspace{1cm}

\bibliographystyle{iopart-num}
\bibliography{Ma_base_de_papier}

\providecommand{\newblock}{}
\begin{thebibliography}{10}
\expandafter\ifx\csname url\endcsname\relax
  \def\url#1{{\tt #1}}\fi
\expandafter\ifx\csname urlprefix\endcsname\relax\def\urlprefix{URL }\fi
\providecommand{\eprint}[2][]{\url{#2}}

\bibitem{Polettini2014_vol141}
Polettini M and Esposito M 2014 {\em J. Chem. Phys.\/} {\bf 141} 024117
  \urlprefix\url{http://scitation.aip.org/content/aip/journal/jcp/141/2/10.1063/1.4886396}

\bibitem{Schnakenberg1976_vol48}
Schnakenberg J 1976 {\em Rev. Mod. Phys.\/} {\bf 48}(4) 571--585
  \urlprefix\url{http://link.aps.org/doi/10.1103/RevModPhys.48.571}

\bibitem{Polettini2016_vol94}
Polettini M, Bulnes-Cuetara G and Esposito M 2016 {\em Phys. Rev. E\/} {\bf
  94}(5) 052117
  \urlprefix\url{http://link.aps.org/doi/10.1103/PhysRevE.94.052117}

\bibitem{Notes_Beijeren2011}
van Beijeren H 2011 Statistical physics of systems out of equilibrium
  \urlprefix\url{http://www.staff.science.uu.nl/~beije101/outofeqseoul.htm}

\bibitem{Broeck2013_vol}
Van~den Broeck C 2013 {\em Proceedings of the International School of Physics
  "Enrico Fermi", Course CLXXXIV Physics of Complex Colloids, C. Bechinger, F.
  Sciortino and P. Ziherl eds., Italian Physical Society\/}
  \urlprefix\url{http://doi.org/10.3254/978-1-61499-278-3-155}

\bibitem{Book_Sekimoto2010}
Sekimoto K 2010 {\em Stochastic Energetics\/} ({\em Lecture Notes in Physics\/}
  vol 799) (Springer, Berlin/Heidelberg)
  \urlprefix\url{http://doi.org/10.1007/978-3-642-05411-2}

\bibitem{Wachtel2015_vol92}
{Wachtel} A, {Vollmer} J and {Altaner} B 2015 {\em Phys. Rev. E\/} {\bf 92}(4)
  042132 \urlprefix\url{http://link.aps.org/doi/10.1103/PhysRevE.92.042132}

\bibitem{Garrahan2016_vol2016}
{Garrahan} J~P 2016 {\em J. Stat. Mech: Theory Exp.\/} {\bf 2016} 073208
  \urlprefix\url{http://dx.doi.org/10.1088/1742-5468/2016/07/073208}

\bibitem{Evans2005_vol38}
Evans R~M~L 2005 {\em J. Phys. A: Math. Gen.\/} {\bf 38} 293
  \urlprefix\url{http://stacks.iop.org/0305-4470/38/i=2/a=001}

\bibitem{Popkov2010_vol2010}
Popkov V, Schuetz G~M and Simon D 2010 {\em J. Stat. Mech: Theory Exp.\/} {\bf
  2010} P10007
  \urlprefix\url{http://stacks.iop.org/1742-5468/2010/i=10/a=P10007}

\bibitem{Chetrite2013_vol111}
Ch{\'e}trite R and Touchette H 2013 {\em Phys. Rev. Lett.\/} {\bf 111}(12)
  120601 \urlprefix\url{http://link.aps.org/doi/10.1103/PhysRevLett.111.120601}

\bibitem{Budini2014_vol2014}
Budini A~A, Turner R~M and Garrahan J~P 2014 {\em J. Stat. Mech: Theory Exp.\/}
  {\bf 2014} P03012
  \urlprefix\url{http://stacks.iop.org/1742-5468/2014/i=3/a=P03012}

\bibitem{Chetrite2015_vol16}
Ch{\'e}trite R and Touchette H 2015 {\em Annales Henri Poincar{\'e}\/} {\bf 16}
  2005--2057 \urlprefix\url{http://dx.doi.org/10.1007/s00023-014-0375-8}

\bibitem{Chetrite2015_vol2015}
Ch{\'e}trite R and Touchette T 2015 {\em J. Stat. Mech: Theory Exp.\/} {\bf
  2015} P12001
  \urlprefix\url{http://stacks.iop.org/1742-5468/2015/i=12/a=P12001}

\bibitem{Chabane_2020}
Chabane L, Ch{\'e}trite R and Verley G 2020 {\em Journal of Statistical
  Mechanics: Theory and Experiment\/} {\bf 2020} 033208
  \urlprefix\url{http://pperso.th.u-psud.fr/page_perso/Verley/Papers/Chabane_2020vol.pdf}

\bibitem{Touchette2017_vola}
{Touchette} H 2017 {\em arXiv\/} (\textit{Preprint} \eprint{1708.02890})
  \urlprefix\url{http://adsabs.harvard.edu/abs/2017arXiv170802890T}

\bibitem{Zia2007_vol}
Zia R~K~P and Schmittmann B 2007 {\em J. Stat. Mech: Theory Exp.\/}
  \urlprefix\url{http://stacks.iop.org/1742-5468/2007/i=07/a=P07012}

\bibitem{Polettini2012_vol97}
Polettini M 2012 {\em Europhys. Lett.\/} {\bf 97} 30003
  \urlprefix\url{http://stacks.iop.org/0295-5075/97/i=3/a=30003}

\bibitem{Andrieux2012_vola}
{Andrieux} D 2012 {\em arXiv\/} (\textit{Preprint} \eprint{1208.5699})
  \urlprefix\url{http://arxiv.org/abs/1208.5699}

\bibitem{Andrieux2012_vol}
Andrieux D 2012 {\em arXiv\/} (\textit{Preprint} \eprint{1212.1807})
  \urlprefix\url{http://arxiv.org/abs/1212.1807}

\bibitem{Verley2016_vol93}
Verley G 2016 {\em Phys. Rev. E\/} {\bf 93}(1) 012111
  \urlprefix\url{http://pperso.th.u-psud.fr/page_perso/Verley/Papers/Verley2016_vol93.pdf}

\bibitem{Rao2018vol149}
Rao R and Esposito M 2018 {\em The Journal of chemical physics\/} {\bf 149}
  245101 \urlprefix\url{https://doi.org/10.1063/1.5042253}

\bibitem{Andrieux2007_vol127}
Andrieux D and Gaspard P 2007 {\em J. Stat. Phys.\/} {\bf 127} 107--131

\bibitem{Altaner2015_vol92}
Altaner B, Wachtel A and Vollmer J 2015 {\em Phys. Rev. E\/} {\bf 92}(4) 042133
  \urlprefix\url{http://link.aps.org/doi/10.1103/PhysRevE.92.042133}

\bibitem{Jack2010_vol184}
Jack R~L and Sollich P 2010 {\em Prog. Theor. Phys. Suppl.\/} {\bf 184}
  304--317
  \urlprefix\url{http://ptps.oxfordjournals.org/content/184/304.abstract}

\bibitem{Chabane2021_vol}
Chabane L, Lazarescu A and Verley G 2021  (\textit{Preprint}
  \eprint{2109.06830})
  \urlprefix\url{http://pperso.th.u-psud.fr/page_perso/Verley/Papers/Chabane_2021vol.pdf}

\bibitem{Maes2020vol850}
Maes C 2020 {\em Physics Reports\/} {\bf 850} 1--33
  \urlprefix\url{http://www.sciencedirect.com/science/article/pii/S0370157320300120}

\bibitem{Lau2007_vol99}
Lau A~W~C, Lacoste D and Mallick K 2007 {\em Phys. Rev. Lett.\/} {\bf 99}(15)
  158102 \urlprefix\url{http://link.aps.org/doi/10.1103/PhysRevLett.99.158102}

\bibitem{Lacoste2008_vol78}
Lacoste D, Lau A~W and Mallick K 2008 {\em Phys. Rev. E\/} {\bf 78} 011915

\bibitem{Vroylandt2018vol2018}
Vroylandt H, Lacoste D and Verley G 2018 {\em J. Stat. Mech: Theory Exp.\/}

\end{thebibliography}

\end{document}